\documentclass[a4paper, 12pt, one column]{article}
\usepackage{graphicx}
\begin{document}
\title{ACCRETION, PRIMORDIAL BLACK HOLES AND STANDARD COSMOLOGY}
\author{B. Nayak$^1$ and L. P. Singh$^2$\\Department of Physics\\ Utkal University\\ Bhubaneswar-751004\\ India\\$^1$bibeka@iopb.res.in\\$^2$lambodar\_uu@yahoo.co.in\\ }
\date{ }
\maketitle
\begin{abstract}
Primordial Black Holes evaporate due to Hawking radiation. We find that the evaporation times of primordial black holes increase when accretion of radiation is included. Thus, depending on accretion efficiency, more and more number of primordial black holes are existing today, which strengthens the conjecture that the primordial black holes are the proper candidate for dark matter. 
\end{abstract}


\newpage
\section{INTRODUCTION}
Black holes which are formed in the early Universe are known as Primordial Black Holes (PBHs). A comparison of the cosmological density of the Universe at any time after the Big Bang with the density associated with a black hole shows that PBHs would have of order the particle horizon mass. PBHs could thus span enormous mass range starting from $10^{-5}gm$ to more than $10^{15}gm$. These black holes are formed as a result of initial inhomogeneities[1,2], inflation[3,4], phase transitions[5], bubble collisions[6,7] or the decay of cosmic loops[8]. In 1974 Hawking discovered that the black holes emit thermal radiation due to quantum effects[9]. So the black holes get evaporated depending upon their masses. Smaller the masses of the PBHs, quicker they evaporate. But the density of a black hole varies as inversely with its mass. So high density is needed for forming lighter black holes. And such high densities is available only in the early Universe. Thus Primordial Black Holes are the only black holes whose masses could be small enough to have evaporated by present time. Further, PBHs could act as seeds for structure formation[10] and could also form a significant component of dark matter[11,12,13].

Since the cosmological enviornment is very hot and dense
in the radiation-dominated era, it is expected that appreciable absorption of
the energy-matter from the surroundings could take place.
Calculation of such PBH accretion in standard cosmology have a long
history but are plauged
with significant uncertainties. The early work by Zel'dovich and Novikov[1]
speculated that PBHs might even be able to grow as fast as the horizon.
Subsequent works, especially by Carr and Hawking[2,14], made a convincing case
that such growth could not occur and moreover that once the PBH became
significantly smaller than the horizon,
accretion would become very inefficient.
But it  has
been noticed that such accretion is most effective in altered gravity
scenarios.
This accretion is responsible for the prolongation of the lifetime of PBHs in
braneworld models[15] as well as in scalar-tensor models[16,17].

Using standard cosmology Barrow and Carr[18] have studied the evaporation of PBHs. They have, however, not included the effect of accretion of radiation which seems to play an important role in scalar-tensor models. Majumdar, Das Gupta and Saxena[19] have provided a viable solution of the baryon asymmetry problem including accretion. In the  present work, we include accretion of radiation while studying the evaporation of PBHs and have shown that how evaporation times of PBHs change with accretion efficiency.
\section{PBH EVAPORATION IN STANDARD COSMOLOGY}
For a spatially flat(k=0) FRW Universe with scale factor $a$, the Einstein equation is[20]
\begin{eqnarray}
\Big(\frac{\dot{a}}{a}\Big)^2 =\frac{8\pi G}{3} \rho
\end{eqnarray}
where $\rho$ is the density of the Universe.\\
The energy conservation equation is
\begin{eqnarray}
\dot{\rho} + 3 \Big(\frac{\dot{a}}{a}\Big)({1+\gamma})\rho=0
\end{eqnarray}
on assuming that the universe is filled with perfect fluid describrd by equation of state $p=\gamma \rho$ . The parameter $\gamma$ is $\frac{1}{3}$ for radiation dominated era$(t<t_1)$ and is $0$ for matter dominated era$(t>t_1)$, where time $t_1$ marks the end of the radiation dominated era $\approx 10^{11}$ sec.\\
Now equation(2) gives
\begin{eqnarray*}
\rho \propto \left\{
\begin{array}{rr}
a^{-4} &  (t<t_1)\\
a^{-3} &  (t>t_1)
\end{array}
\right.
\end{eqnarray*}
Using this solution in equation (1), one gets the wellknown temporal behaviour of the scale factor $a(t)$ as
\begin{eqnarray}
a(t) \propto \left\{
\begin{array} {rr}
t^{\frac{1}{2}} &  (t<t_1)\\
t^{\frac{2}{3}} &  (t>t_1)
\end{array}
\right.
\end{eqnarray}

Due to Hawking evaporation, the rate at which the PBH mass (M) decreases is given by
\begin{eqnarray}
\dot{M}_{evap}=-4\pi r_{BH}^2 a_H T_{BH}^4
\end{eqnarray}
where $r_{BH} \sim$ black hole radius=$2GM$ with G as Newton's gravitational constant.\\
$~~~~~~~~~~ a_H \sim$ black body constant\\
$~~$ and$~~ T_{BH} \sim$ Hawking Temperature=$\frac{1}{8\pi GM}$ .\\
Now equation (4) becomes 
\begin{eqnarray}
\dot{M}_{evap}=-\frac{a_H}{256\pi^3} \frac{1}{G^2 M^2}
\end{eqnarray}
Integrating the above equation, we get
\begin{eqnarray}
M=\Big[M_i^3 + 3 \alpha (t_i-t)\Big]^{\frac{1}{3}}
\end{eqnarray}
where $\alpha=\frac{a_H}{256 \pi^3}\frac{1}{G^2}$ and $M_i$ is the black hole mass at its formation time $t_i$ . It is worthwhile to remark that we assume $M_i$ to be same as the horizon mass as conjectured in [21]. We will, however, demonstrate in the following that two masses will have different temporal growth.
\section{ACCRETION}
When a PBH passes through radiation dominated era, the accretion of radiation leads to increase of its mass with the rate given by
\begin{eqnarray}
\dot{M}_{acc}=4\pi fr_{BH}^2 \rho_r
\end{eqnarray}
where $\rho_r$ is the radiation energy density of the sorrounding of the black hole=$\frac{3}{8\pi G} \Big(\frac{\dot{a}}{a}\Big)^2$ and $f$ is the accretion efficiency. The value of the accretion efficiency $f$ depends upon complex physical processes such as the mean free paths of the particles comprising the radiation sorrounding the PBHs. Any peculiar velocity of the PBH with respect to the cosmic frame could increase the value of $f$[19,22]. Since the precise value of $f$ is unknown, it is customary[23] to take the accretion rate to be proportional to the product of the surface area of the PBH and the energy density of radiation with $f \sim O(1)$.  \\
After substituting the expressions for $r_{BH}$ and $\rho_R$ equation(7) becomes
\begin{eqnarray}
\dot{M}_{acc}=6fG\Big(\frac{\dot{a}}{a}\Big)^2 M^2
\end{eqnarray}
Using equation(3), we get
\begin{eqnarray}
\dot{M}_{acc}=\frac{3}{2} f G \frac{M^2}{t^2}
\end{eqnarray}
On integration, the above eqution gives
\begin{eqnarray}
M(t)=\Big[M_i^{-1} +\frac{3}{2}fG\Big(\frac{1}{t}-\frac{1}{t_i}\Big)\Big]^{-1}
\end{eqnarray}

Using horizon mass which varies with time as $M_H(t)=G^{-1}t$, as initial mass of PBH, we get
\begin{eqnarray}
M(t)=M_i\Big[1 +\frac{3}{2}f\Big(\frac{t_i}{t}-1\Big)\Big]^{-1}
\end{eqnarray}
We draw two important conclusions from equation (11).

First we obtain the variation of accreting mass with time for different $f$ as shown in Figure-1. The figure clearly indicates that the mass of the PBH increases with accretion efficiency. 
\begin{figure}[h]
\centering
\includegraphics[scale=0.5]{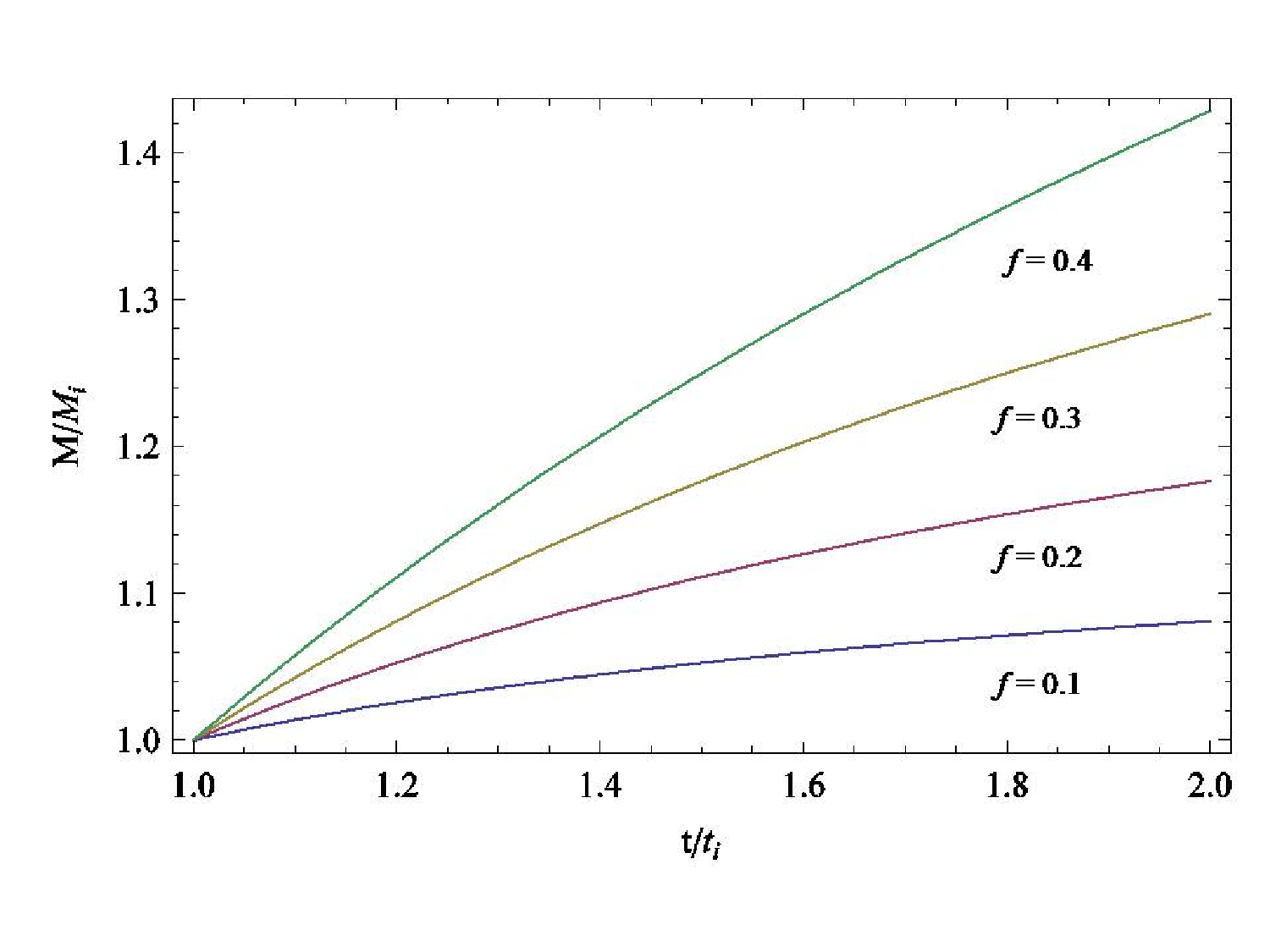}
\caption{Variation of accreting mass for $f=0.1,0.2,0.3,0.4$}
\label{fig1}
\end{figure}

For large $t$, $M_{BH}$ of equation (11) asymptotes to its maximum value as
 \begin{eqnarray}
M_{max}=\frac{M_i}{1-\frac{3}{2}f}
\end{eqnarray}
which leads to an upperbound, 
\begin{eqnarray}
f<\frac{2}{3}
\end{eqnarray}

The second conclusion is with regard to variation of PBH mass visavis that of horizon with time. 
\begin{figure}[h]
\centering
\includegraphics[scale=0.5]{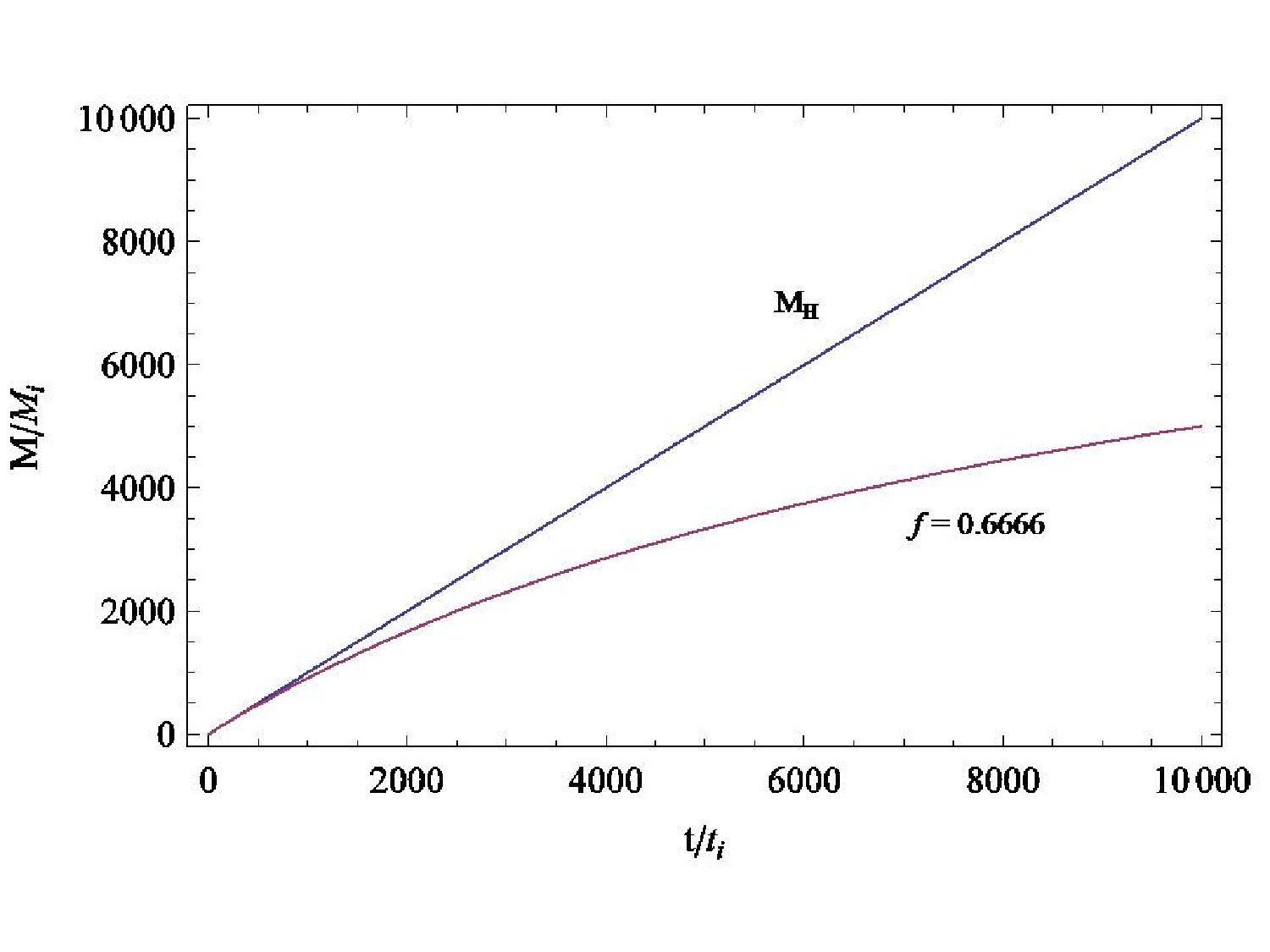}
\caption{Variation of $M_{PBH}$ having $f=0.6666$ and $M_H$ with $t$}
\label{fig2}
\end{figure}

Since the horizon mass grows as $M_H(t)\sim G^{-1}t$, from equation (11) one finds that $M_H$ grows faster than the black hole mass $M_{BH}$. This is graphically shown in Figure-2. Thus enough radiation density is
available within the cosmological horizon for a PBH to accrete causally, making
accretion effective in this scenario.
\section{PBH DYNAMICS IN DIFFERENT ERA}
Primordial Black Holes, as discussed before, are only formed in radiation dominated era. So depending on their evaporation, we can divide PBHs into 2 categories.\\
(i) PBHs evaporating in radiation dominated era $(t<t_1)$\\
(ii) PBHs evaporating in matter dominated era $(t>t_1)$.\\
\textbf{CASE-I}  $(t<t_1)$\\
Black hole evaporation equation (6) implies
\begin{eqnarray}
M=M_i\Big[1+\frac{3 \alpha}{M_i^3}(t_i-t) \Big]^\frac{1}{3}
\end{eqnarray}
If we consider both evaporation and accretion simultaneously, then the rate at which primordial black hole mass changes is given by
\begin{eqnarray}
\dot{M}_{PBH}=\frac{3}{2}fG \frac{M^2}{t^2}-\alpha \frac{1}{M^2}
\end{eqnarray}
This equation can not be solved analytically. So we have solved it by using numerical methods.\\

For PBHs with formation mass $M_i^2 > \frac{a_H}{384 f G}$, the
magnitude of the first term (accretion) exceeds that of the second term
(evaporation).
In the radiation dominated era for a PBH whose formation mass satisfies
the above relation, accretion is dominant upto a value of $t$, say $t_{c}$, at which accretion rate equals evaporation rate (the PBH mass
rises to a maximum value $M_{max}$ at this stage), and after that evaporation dominates over accretion.
 For our calculation purpose, we have used $\alpha \approx G^{-2}=10^{28}(\frac{gm^3}{sec})$ and $G=10^{-38}(\frac{sec}{gm})$.\\
For a given $M_i$, the solution as given by equation(14) and the solution of the equation(15) are shown in Figure-3. The figure clearly shows that the evaporation time of PBH increases with accretion efficiency.
\begin{figure}[h]
\centering
\includegraphics[scale=0.5]{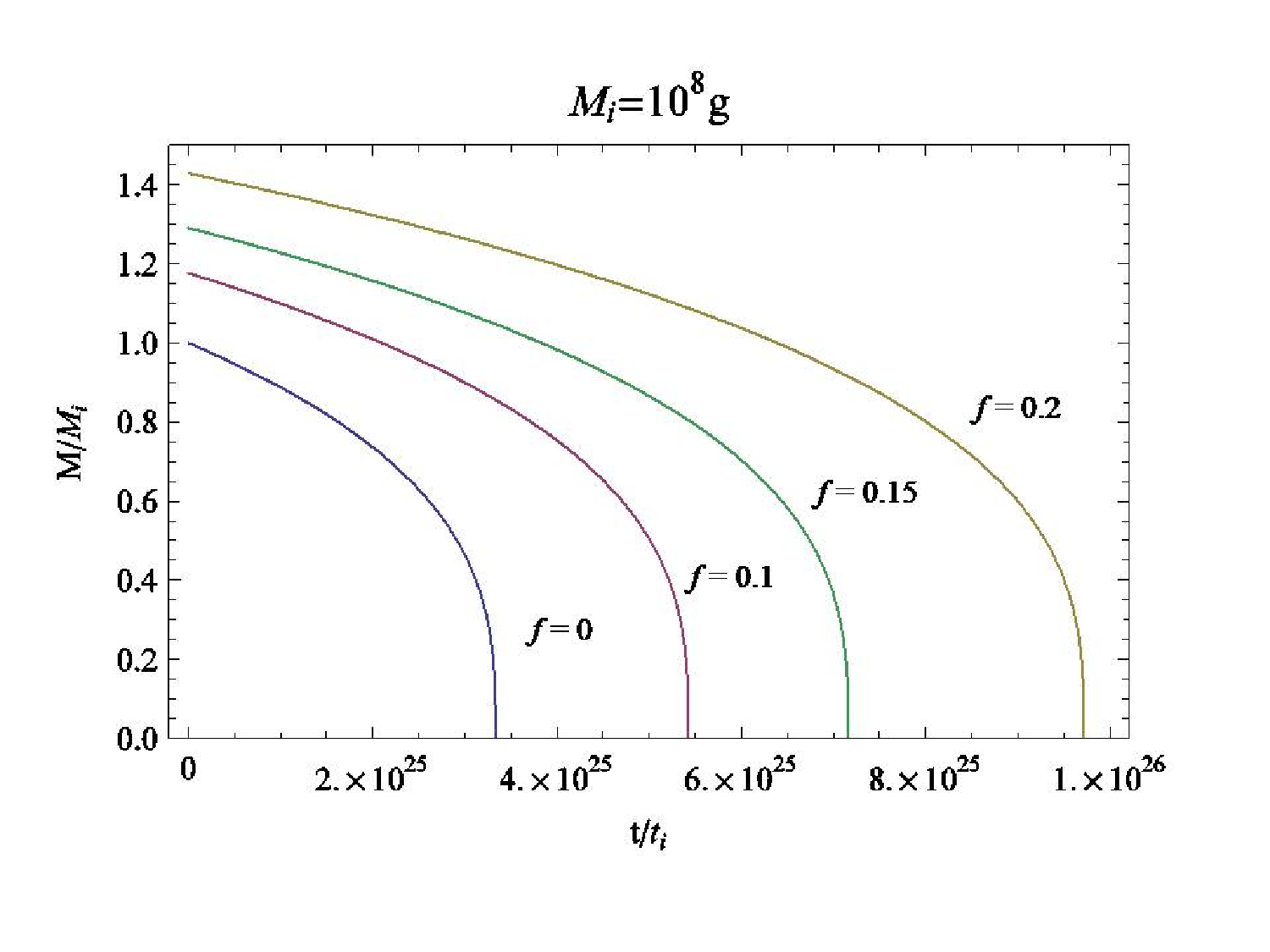}
\caption{Variation of PBH mass for $f=0,0.1,0.15,0.2$}
\label{fig3}
\end{figure}

\textbf{CASE-II}  $(t>t_1)$\\
Since there is no accretion in matter dominated era, so the first term in the combined equation (15) for variation of $M_{PBH}$ with time needs to be integrated only upto $t_1$ .\\
Based on numerical solution with above provision, we construct the Table-1 for the PBHs which are evaporating at present time.\\
\begin{table}
\begin{tabular}[c]{|c|c|c|}
\hline
\multicolumn{3}{|c|}{$t_{evap}=t_0=4.42 \times10^{17} s$}\\
\hline
$f$  &  $t_i$  &  $M_i$  \\
\hline
$0$  &  $2.3669\times10^{-23}$s  &  $2.3669\times10^{15}$g  \\
\hline
$0.1$  &  $2.0119\times10^{-23}$s  &  $2.0119\times10^{15}$g  \\
\hline
$0.2$  &  $1.6568\times10^{-23}$s  &  $1.6568\times10^{15}$g  \\
\hline
$0.3$  &  $1.3018\times10^{-23}$s  &  $1.3018\times10^{15}$g  \\
\hline
$0.4$  &  $0.9467\times10^{-23}$s  &  $0.9467\times10^{15}$g  \\
\hline
$0.5$  &  $0.5916\times10^{-23}$s  &  $0.5916\times10^{15}$g  \\
\hline
$0.6$  &  $0.23669\times10^{-23}$s  &  $0.23669\times10^{15}$g  \\
\hline
\end{tabular}
\caption{The formation times and initial masses of the PBHs which are evaporating now are displayed for several
accretion efficiencies.}
\end{table}

It is clear from the table that accretion makes it possible for the PBHs evaporating now to be formed at earlier times with smaller initial masses.
\section{CONSTRAINTS ON PBH}
The fraction of the Universes' mass going into PBHs at time $t$ is[2]
\begin{eqnarray}
\beta(t)=\Big[\frac{\Omega_{PBH}(t)}{\Omega_R}\Big](1+z)^{-1}
\end{eqnarray}
where $\Omega_{PBH}(t)$ is the density parameter associated with PBHs formed at time $t$, $z$ is the redshift associated with time $t$. $\Omega_R$ is the microwave background density having value $10^{-4}$.\\
For $t<t_1$, redshift defination implies, $(1+z)^{-1}=\Big(\frac{t}{t_1}\Big)^{\frac{1}{2}} \Big(\frac{t_1}{t_0}\Big)^{\frac{2}{3}}$ . \\
So
\begin{eqnarray}
\beta(t)=\Big(\frac{t}{t_1}\Big)^{\frac{1}{2}} \Big(\frac{t_1}{t_0}\Big)^{\frac{2}{3}} \Omega_{PBH}(t) \times 10^4
\end{eqnarray}
Using $M=G^{-1}t$, we can transcribe the equation (17) to write the fraction of the Universe going into PBHs' as a function of mass M is
\begin{eqnarray}
\beta(M)=\Big(\frac{M}{M_1}\Big)^{\frac{1}{2}} \Big(\frac{t_1}{t_0}\Big)^{\frac{2}{3}}\Omega_{PBH}(M) \times 10^4
\end{eqnarray}
Observations of the cosmolgical deceleration parameter imply $\Omega_{PBH}(M)<1$ over all mass ranges for which PBHs have not evaporated yet. But presently evaporating PBHs($M_*$) generate a $\gamma$-ray background whose most of the energy is appearing at around 100 Mev[24]. If the fraction of the emitted energy which goes into photons is $\epsilon_{\gamma}$, then the density of the radiation at this energy is expected to be $\Omega_{\gamma}=\epsilon_{\gamma}\Omega_{PBH}(M_*)$.
Since $\epsilon_{\gamma}\sim 0.1$ and the observed $\gamma$-ray background density around $100$ Mev is $\Omega_{\gamma} \sim 10^{-9}$, one gets $\Omega_{PBH}<10^{-8}$ . \\
Now equation (18),therefore, becomes
\begin{eqnarray}
\beta(M_*) < \Big(\frac{M_*}{M_1}\Big)^{\frac{1}{2}} \times \Big(\frac{t_1}{t_0}\Big)^{\frac{2}{3}} \times 10^{-4}
\end{eqnarray}
The variation of $\beta(M_*)$ with $f$ drawn from variation of $M_*$ with $f$ is shown in the Table-2. The bound on $\beta(M_*)$ is strengthened as $f$ approaches its maximum value $2/3$.

\begin{table}
\begin{tabular}[c]{|c|c|c|}
\hline
\multicolumn{3}{|c|}{$t_{evap}=t_0$}\\
\hline
$f$  &  $M_*$  &  $\beta(M_*) <$\\
\hline
$0$  &  $2.3669 \times 10^{15}$g  &  $5.71227 \times 10^{-26}$\\
\hline
$0.2$  &  $1.6568 \times 10^{15}$g  &  $4.77918 \times 10^{-26}$\\
\hline
$0.4$  &  $9.467 \times 10^{14}$g  &  $3.61264 \times 10^{-26}$\\
\hline
$0.6$  &  $2.3669 \times 10^{14}$g  &  $1.80638 \times 10^{-26}$\\
\hline
$0.666$  &  $2.36689 \times 10^{12}$g  &  $1.80637 \times 10^{-27}$\\
\hline
$0.6666$  &  $2.36687 \times 10^{11}$g  &  $5.71233 \times 10^{-28}$\\
\hline
\end{tabular}
\caption{Upper bounds on the initial mass fraction of PBHs that are
evaporating today for various accretion efficiencies $f$.}
\end{table}

\section{DISCUSSION AND CONCLUSION}
Consideration of evaporation alone makes the Primordial Black Holes which are created on or before $2.3669 \times 10^{-23}~ sec$ completely evaporate by the present time. However, we found that if we include accretion, then the Primotdial Black Holes which are created at the same instant of time will live longer depending on their accretion efficiency. Our analysis also leads to an upperbound on the accretion efficiency as $f<\frac{2}{3}$. Further, the constraint on the fraction of the Universes' mass going into PBHs' obtained by us is consistent with previous results[25,26] that $\beta(M_*)<10^{-25}$.

Thus accretion increases the number of existing PBHs depending on accretion efficiency, which lends support to the proposal of considering PBHs as the viable candidate for dark matter. We, thus, provide within standard cosmology a possible realisation of the speculation advanced earlier[11,12,13].

In the present context, one may consider back reaction of primordial black hole evaporation which can lead to non-trivial consequences[27]. Back reaction modifies the radius and temperature of PBH [28] which ultimately affects the accretion and evaporation rates. Thus it might be interesting to see in what way resulting modification could in turn impact the evolution of black holes. Such effects, it is argued [29], may make the Hawking process terminate while the PBH still has macroscopic mass. There are also competing speculations that blackholes completely evaporate leaving no remnants [30] or that blackholes cease to evaporate as they approach Planck mass [31]. Whatever may be the cause of the stability of final remnant of radiating PBHs, the finite mass relics would provide a possible cold dark matter candidate [32].
\section*{ACKNOWLEDGEMENT}
We are thankful to Institute of Physics, Bhubaneswar, India, for providing the library and computational facility. B.Nayak would like to thank the Council of Scientific and Industrial Research, Government of India, for the award of SRF, F.No. 09/173(0125)/2007-EMR-I .
\section*{REFERENCES}
$[1]$ Ya. B. Zeldovich and I. Novikov, Sov. Astron. Astrophys. J. $\textbf{10}$, $602$ $(1967)$ .
$[2]$ B. J. Carr, Astrophys. J. $\textbf{201}$, $1$ $(1975)$ .\\
$[3]$ M. Y. Kholpov, B. A. Malomed and Ya. B. Zeldovich, Mon. Not. R. Astron. 

Soc. $\textbf{215}$, $575$ $(1985)$ .\\
$[4]$ B. J. Carr, J. Gilbert and J. Lidsey, Phys. Rev. D $\textbf{50}$ $4853$ $(1994)$ . \\
$[5]$ M. Y. Kholpov and A. Polnarev, Phys. Lett. $\textbf{97B}$, $383$ $(1980)$ . \\
$[6]$ H. Kodma, M. Sasaki and K. Sato, Prog. Theor. Phys. $\textbf{68}$, $1079$ $(1982)$ .\\
$[7]$ D. La and P. J. Steinhardt, Phys. Rev. Lett $\textbf{62}$, $376$ $(1989)$ . \\
$[8]$ A. Polnarev and R. Zemboricz, Phys. Rev. D $\textbf{43}$, $1106$ $(1988)$ . \\
$[9]$ S. W. Hawking, Commun. Math. Phys. $\textbf{43}$, $199$ $(1975)$ .\\
$[10]$ K. J. Mack, J. P. Ostriker and M. Ricotti, Astrophys. J. $\textbf{665}$, $1277$ $(2007)$ .\\
$[11]$ D. Blais, C. Kiefer, D. Polarski, Phys. Lett. B $\textbf{535}$, $11$ $(2002)$ .\\
$[12]$ D. Blais, T. Bringmann, C. Kiefer, D. Polarski, Phys. Rev. D $\textbf{67}$, $024024$ 

$(2003)$.\\
$[13]$ A. Barrau, D. Blais, G. Boudoul, D. Polarski, Annalen Phys. $\textbf{13}$, $114$ $(2004)$.\\ 
$[14]$ B. J. Carr and S. W. Hawking, Mon. Not. R. Astron. Soc. $\textbf{168}$, $399$ $(1974)$.\\
$[15]$ A. S. Majumdar, Phys. Rev. Lett. $\textbf{90}$, $031303$ $(2003)$.\\
$[16]$ R. Guedens, D. Clancy  and A. R. Liddle, Phys. Rev. D $\textbf{66}$, $083509$ $(2002)$ ; 

A. S. Majumdar and N. Mukherjee, Int. J. Mod. Phys. D $\textbf{14}$, $1095$ $(2005)$ ;

A. S. Majumdar, D. Gangopadhyay and L. P. Singh, Mon. Not. R. Astron. 

Soc. $\textbf{385}$, $1467$ $(2008)$.\\
$[17]$ B. Nayak, L. P. Singh and A. S. Majumdar, Phys. Rev. D $\textbf{80}$, 023529 (2009).\\
$[18]$ J. D. Barrow and B. J. Carr, Phys. Rev. D $\textbf{54}$, $3920$ $(1996)$ .\\
$[19]$ A. S. Majumdar, P. Das Gupta and R. P. Saxena, Int. J. Mod. Phys. D $\textbf{4}$, $517$ 

$(1995)$ .\\
$[20]$ S. Wienberg, `Gravitation and Cosmology', Wiley, New York, $1972$ .\\
$[21]$ B. J. Carr, *Erice 2000, Phase transitions in the early universe* 451-469 ; 

B. J. Carr, Lect. Notes Phys. $\textbf{631}$, $301$ $(2003)$.\\
$[21]$ N. Upadhyay, P. Das Gupta and R. P. Saxena, Phys. Rev. D $\textbf{60}$, $063513$ $(1999)$.\\
$[22]$ R. Guedens, D. Clancy and A. R. Liddle, Phys. Rev. D $\textbf{66}$, $083509$ $(2002)$ .\\
$[23]$ D. Page and S. W. Hawking, Astrophys. J. $\textbf{206}$, $1$ $(1976)$ .\\
$[24]$ B. J. Carr, Astron. Astrophys. Trans. $\textbf{5}$, $43$ $(1994)$ .\\
$[25]$ I. D. Novikov et al., Astron. Astrophys. J. $\textbf{80}$, $104$ $(1979)$ .\\
$[26]$ J. MacGibbon and B. J. Carr, Astrophys. J. $\textbf{371}$, $447$ $(1991)$ .\\
$[27]$ T. Buchert, Gen. Rel. Grav. $\textbf{40}$, $467$ $(2008)$ ;
G. F. R. Ellis, Nature $\textbf{452}$, $158$ 

$(2008)$ ;
E. W. Kolb, V. Marra and S. Matarrese, Phys. Rev. D $\textbf{78}$, $103002$ 

$(2008)$.\\
$[28]$ C. O. Lousto and N. Sanchez, Phys. Lett. B $\textbf{212}$, $411$ $(1988)$.\\
$[29]$ L. Susskind and L. Thorlacius, Nucl. Phys. B $\textbf{382}$, $123$ $(1992)$.\\
$[30]$ S. W. Hawking, Phys. Rev. D $\textbf{14}$, $2460$, $(1976)$.\\
$[31]$ K. Maeda, Class. Quant. grav. $\textbf{3}$, $233$, $(1986)$ ; M. J. Bowick et al., Phys. Rev. 

Lett. $\textbf{61}$, $2823$, $(1988)$ ; S. Coleman, J. Preskill and F. Wilczek, Mod. Phys. 

Lett. A $\textbf{6}$, $1631$, $(1991)$. \\
$[32]$ J. H. MacGibbon, Nature $\textbf{329}$, $308$, $(1987)$.\\
\end{document}